# Light guiding and optical resonances in ZnS microstructures doped with Ga or In

B. Sotillo,[a,b,*] P. Fernández[a,*] and J. Piqueras[a]

The high refractive index of ZnS (about 2.4 in the visible range) makes ZnS microstructures good candidates for the fabrication of light guides and resonant microcavities. Light guiding behaviour has been studied in several ZnS structures. In particular, in this work the behavior of rods and plates has been investigated and it is has been shown that the morphology of the studied structures plays a determinant role on the supported resonant modes. In ZnS:Ga and ZnS:In structures investigated in the present work Fabry-Pérot as well as Whispering Gallery resonant modes are supported. A good agreement has been found between the experimental results and the theoretical model used to describe both types of resonances, which have been used to estimate the refractive index of the doped material. In ZnS:Ga wires with triangular cross-section a lasing effect has been observed in the blue-green range. To our knowledge, it is the first time that resonant modes in visible region have been studied in doped ZnS.

## A Introduction

Micro- and nanostructures of ZnS have attracted much attention in the last years due to their potential applications in field-emission devices[1,2,3], photocatalysis[4,5,6], sensing[7,8,9], electroluminescent devices[10] or nanogenerators[11]. ZnS is a II-VI semiconductor with a direct bandgap in the UV range (3.72 eV for zinc-blende phase and 3.77 eV for wurtzite phase), a high exciton binding energy (about 40 meV) and good chemical stability, among other properties.

Properties and potential applications of pure ZnS micro- and nanostructures have been widely investigated, but fewer studies have been published about doped structures[12]. Common dopants in ZnS are for example Mn[13,14], Cu[15,16], Ni[17,18] or Eu[19,20]. The incorporation of impurities (activators and coactivators) into ZnS has been traditionally performed to alter the luminescence properties of bulk material[21]. In previous works the changes in morphology and luminescence emission associated to the incorporation of III-group dopants (Al, Ga, In) into ZnS micro- and nanostructures grown by thermal evaporation-deposition methods have been studied[22,23,24]. In this work the main focus is on the changes induced by different dopants in the refractive index. Although it has been traditionally less exploited for applications than other properties, the high refractive index, around 2.4 in the visible region, is one of the most interesting properties of ZnS. This high value of refractive index makes ZnS structures good candidates to fabricate light guides and resonant microcavities to be used in the fabrication of lasers[25,26], sensors and biosensors[27,28] or optical filters[29,30]. Despite the promising use of micro- and nanostructures of ZnS as resonant cavities, there have been few reports on this topic and only in undoped material. In particular, Zapien et al.[25] used individual ZnS nanoribbons, grown by hydrogen assisted thermal evaporation as active medium for the resonant cavity (Fabry-Pérot type) and light guide for making a nanolaser in the UV. Ding et al.[26] obtained nanowires by a vapour-liquid-solid mechanism (VLS) and studied the laser phenomenon in the collective emission. These nanowires also act as Fabry-Pérot cavities. In both cases the studies are performed in undoped ZnS structures in the UV range.

In this work we have investigated the changes in the refraction index induced by the presence of dopants, in particular In and Ga in ZnS, and the different resonant modes in the visible range (both Fabry-Pérot and Whispering Gallery modes), determining that the doped structures are suitable for being used as resonant microcavities. To our knowledge, it is the first time that resonant modes in visible region have been studied in doped ZnS.

## B Experimental section

Different starting materials have been used to obtain the ZnS structures doped with Ga or with In. For Ga doped ZnS, the starting mixture is composed by ZnS and $Ga_2S_3$ powders, both with a purity of 99.99%, with an amount of $Ga_2S_3$ of 15 wt.% (which corresponds to 6.2 at.% of Ga in the mixture). For In doped material, two mixtures of ZnS and $In_2S_3$ powders (again with purity of 99.99%) have been used, with 5 and 15 wt. % of $In_2S_3$, respectively (which correspond to 1.51 and 4.66 at. % of In). The mixture was homogenized using a centrifugal ball mill during 5 h. The powder was compacted under a compressive load to form disk-shaped pellets. The dimensions of pellets were about 7 mm in diameter and a thickness of 2 mm.

The structures have been grown by a catalyst free evaporation-solidification method. Thermal treatments have been performed in a horizontal tube furnace in $N_2$ atmosphere to avoid oxidation. The furnace chamber was first evacuated to $10^{-2}$ torr, and then the gas was introduced at constant rate, to maintain a pressure of 700 torr. The starting material is placed in the centre of the tube. Different temperatures and treatment times have been used for Ga or In doped material. For ZnS:Ga, the source material is maintained at 1300 ºC for 2 hours, whereas for ZnS:In the treatments have been performed in two steps, the first one at 1000 ºC during 15 hours, and the second at 1200 ⁰C for 2h. In both cases, the structures have been grown at different points of the furnace alumina tube (i.e. at different deposition temperatures). More details about the thermal treatments have been described elsewhere[31,32].

[a] Dept. Física de Materiales, Faculty of Physics, U. Complutense, Ciudad Universitaria, 28040 Madrid, Spain.
[b] Present address: Istituto di Fotonica e Nanotecnologie, CNR, Piazza Leonardo da Vinci 32, 20133, Milano, Italy
* Corresponding authors: bsotillo@gmail.com and arana@ucm.es

X ray diffraction analysis (XRD) of both, the starting material and the synthesized structures, has been done by means of a Philips X'Pert PRO diffractometer. Secondary electron images were obtained using Leica S440 or FEI Inspect scanning electron microscopes (SEM). X-ray microanalyses have been done with a Bruker AXS Quantax system attached to the LEICA SEM. Micro-photoluminescence (μ-PL) measurements have been carried out in a Horiba JobinYvon LABRAM-HR Raman spectrometer using the 325nm line from a He-Cd laser at room temperature.

## C Results and discussion

After thermal treatment, different ZnS:Ga or ZnS:In structures are obtained on the furnace wall ZnS:Ga microstructures are obtained at a deposition temperature between 1075 and 1150 ⁰C. The XRD measurements showed that all the structures grown in this work (Ga doped and In doped) belong to the wurtzite phase of ZnS (JCPD card no. 36-1450).

As reported in previous works[23,24,31,32], the morphology of the structures is strongly dependent from both dopant and treatment conditions. In the particular case of the systems studied in this work plates and rods of different shapes have been mainly obtained. Among the morphologies obtained the most interesting for studying optical resonant modes seem to be pencils (figure 1a) and wires with triangular cross-section (figure 1b). The pencil-like structures have cross dimensions between 20 and 60 μm and lengths of hundreds of microns. The wires have lengths of tens to hundreds of microns and their cross sections are isosceles or equilateral triangles with sides of about 1 to 10 μm. The incorporation of Ga into the structures has been checked by EDX measurements and reaches at.% values of about 1.7-1.8.

ZnS:In micro- and nanostructures are obtained at two ranges of deposition temperatures: 500 ⁰C - 800 ⁰C and 850 ⁰C - 900 ⁰C. Ribbons and swords (figures 2a and b) are obtained with an initial mixture of 5 wt.% of $In_2S_3$ (1.5 at.% of In) in the higher temperature range. Ribbons have thicknesses between 400 nm and 2 μm, whereas swords show thicknesses between 100 nm and 1μm. The width and length of these structures are in the range of microns and tens of microns respectively. On the other hand, plates (figure 2c) are obtained with an initial mixture of 15 wt.% of $In_2S_3$ (4.7 at.% of In) in the lower temperature range. These plates show mainly hexagonal shape, with typical thicknesses between 100 and 200 nm and widths of microns. EDX measurements show that the degree of In incorporation into the structures depends on the deposition temperature, being higher for the plates (0.4 - 1.5 at.%). For the swords and ribbons, the In content is much lower and, in fact close to the detection limit of the EDX technique. In these structures, the presence of In is confirmed by the cathodoluminescence measurements, described elsewhere[32].

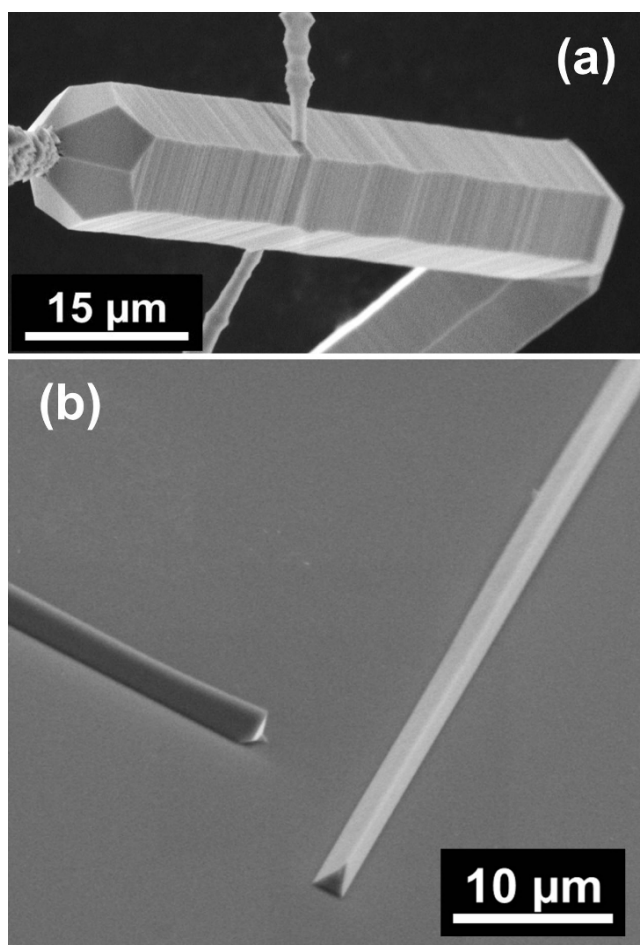

**Figure 1.** SEM images of ZnS:Ga structures: (a) pencil-like structure; (b) Triangular cross-section wires.

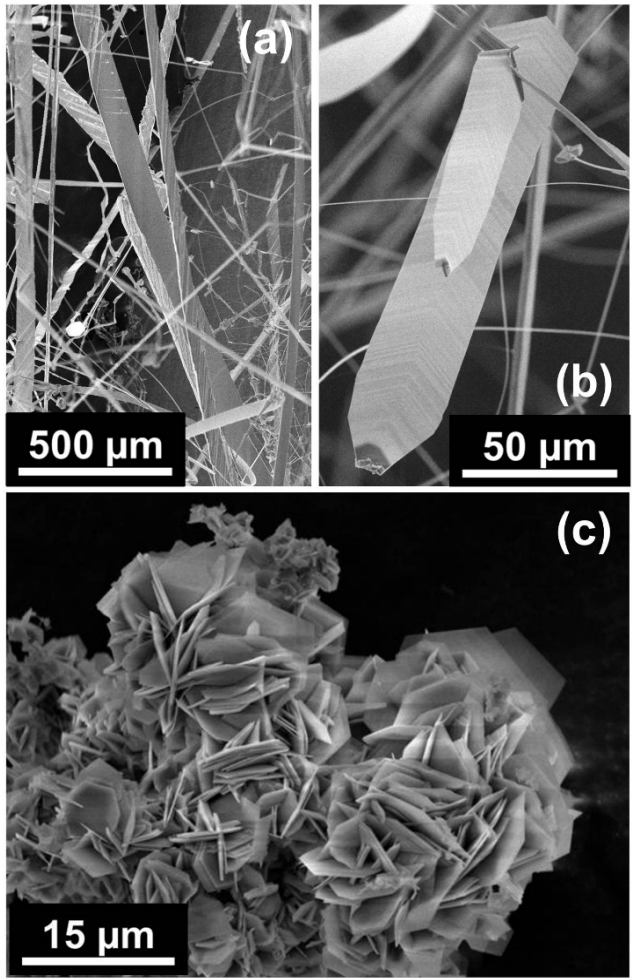

**Figure 2.** SEM images of ZnS:In structures: (a) ribbons; (b) sword-like structures; (c) hexagonal plates.

As mentioned above, ZnS has a refractive index in the visible range (about 2.4 at 500 nm) high enough to make the obtained microstructures good candidates for light guides and resonant microcavities. The behaviour of ZnS structures, in particular swords and wires, as light guides has been tested using both an external light source, and the luminescence emission of the material. In the first configuration one of the sides of the structure is illuminated with the light of a green (532 nm) or a red (650 nm) laser, parallel to the growth axis. In the second configuration, the PL is excited by a laser with peak wavelength at 325 nm. In all the structures (Ga or In doped) studied, efficient light guiding has been observed in both configurations. Some examples are presented in figure 3. In figure 3a the optical image of a ZnS:In sword with a nanowire on the tip is shown, along with the images of the structure illuminated with green and red lasers on the side marked with a black arrow and a star. Intense green or red spots are observed at the tip of the sword opposite to the point of light incidence, which is indicative of the light guiding behaviour. In figure 3b, guiding of PL emission can be observed in ZnS:Ga wires. The yellow spot observed in these images corresponds to the excitation point. Intense blue spots are observed away from the excitation point, showing light guiding over distances larger than 20 μm.

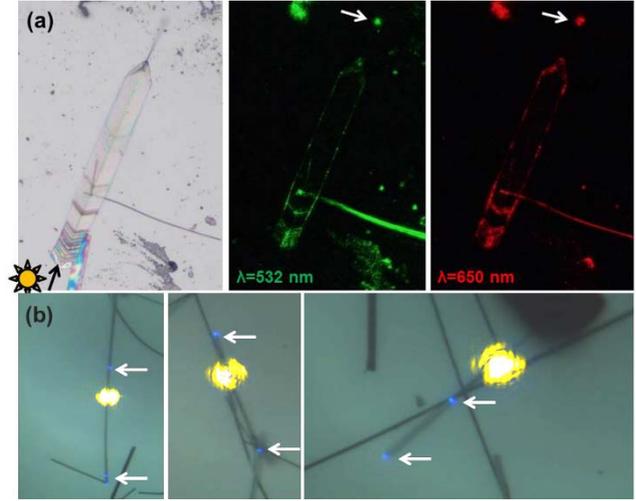

**Figure 3.** (a) Optical image of a ZnS:In sword with a nanowire at the tip, along with the images that show the guiding of green light (532 nm) and red light (650 nm). (b) Optical images of ZnS:Ga wires, where light guiding is observed. The 325 nm incident laser light is the yellow spot, whereas the guided light are the blue spots marked with white arrows.

The different morphologies observed allow studying different optical resonant modes. Fabry-Pérot modes (FPM) have been studied in swords, ribbons and plates of ZnS:In, whereas whispering gallery modes (WGM) have been studied in ZnS:Ga structures as well as in ZnS:In plates.

In figure 4 examples of the PL obtained in swords, ribbons and plates of ZnS:In are presented. In ZnS:In, the PL emission is found mainly in the orange-red region, centred at 610 nm and is ascribed to native and dopant-related defects[32]. In the PL spectra, a clear modulation of the characteristic emission is observed. Due to the geometry of these structures, a possibly origin the observed modulation is the Fabry-Pérot resonances, which are produced when light is confined between two parallel facets of the swords, ribbons or plates. Fabry-Pérot resonant peaks appear at those wavelengths that verify the equation:

$$\lambda = \frac{2 \cdot h \cdot n}{N} \quad (1)$$

where λ is the resonance wavelength, n is the refractive index, N is the interference order and h the length of the FP cavity. h can correspond with the length, the width or the thickness of the structures. The separation between adjacent modes follows the equation:

$$\Delta\lambda = \frac{\lambda^2}{2 \cdot h \cdot \left(n - \lambda \frac{dn}{d\lambda}\right)} \quad (2)$$

Considering n constant (as done for example in ref. 33) the separation is:

$$\Delta\lambda \approx \frac{\lambda^2}{2 \cdot h \cdot n} \quad (3)$$

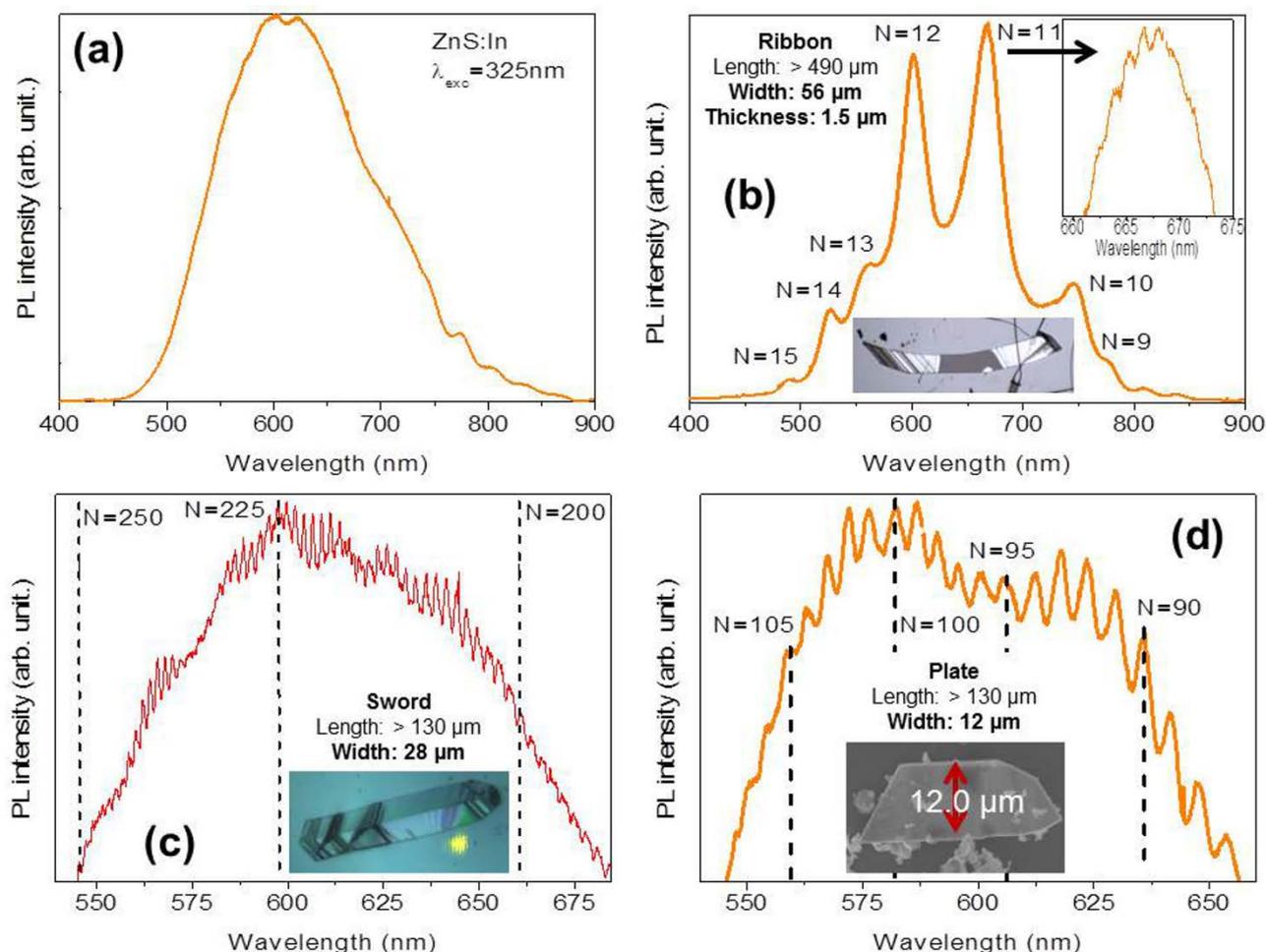

**Figure 4.** (a) Typical PL spectrum of In doped ZnS structures. (b) μ-PL spectrum of a ZnS:In ribbon with the corresponding optical image. The inset shows a detail of the emission maxima corresponding to N=11, where the modulations of the PL associated with the FP modes in the thickness and width of the ribbon are also observed. (c) μ-PL spectrum of a ZnS:In sword and corresponding optical image of the structure. (d) μ-PL spectrum and optical image of a ZnS:In plate. The optical path of the FP mode is marked with a red arrow.

The separation between modes will be higher for smaller FP cavities. Thus for a distance h of 20 microns or less, the maximum spacing would be $\Delta\lambda > 4$ nm, so modes supported in the cavities formed in the width and thickness of ribbons and swords could be observed in the PL measurements. On the contrary, if we consider the long dimension of the ribbon, with an h value of about 100 μm and at emission maximum (620 nm), the difference between interference maxima would be $\Delta\lambda \approx 0.8$ nm, and since our spectral resolution is 1nm, FP modes in the long direction of the ribbon cannot be resolved with the experimental system employed.

In figure 4b, the optical image of a ribbon with a width of 56 μm and a thickness of about 1.5 μm, along with its μ-PL spectrum, is presented. In this case, two types of modulations are observed: the first one, with a higher separation between maxima ($\Delta\lambda \approx 65.5$ nm around 620 nm), can clearly be seen in spectrum of figure 4b (middle), which can be associated, using equation (3.2.3), with resonant modes supported in the thickness of the ribbon. The second modulation, with a smaller separation between maxima ($\Delta\lambda \approx 1.4$ nm around 670 nm), is observed when a selected area around one maximum is studied (figure 4b inset), and it arises from modes supported in the width of the ribbon. Modulation ($\Delta\lambda \approx 2.3$ nm around 620 nm) along the width of 28 μm of a sword can also be seen in the spectra of figure 4c. Finally, FP modes in plates ($\Delta\lambda \approx 5.3$ nm around 620 nm), supported along the dimension marked with a red arrow in figure 4d (12.0 μm), are observed in the PL spectra. Since the refractive index, n, depends also on the wavelength, we have used the dispersion equation for bulk pure material[34]:

$$n^2 = 8.34096 + \frac{0.14540}{\lambda^2 - 0.23979^2} + \frac{3.23924}{\frac{\lambda^2}{36.525^2} - 1} \quad (4)$$

to estimate the n value for undoped material. Then, by means of eqns. (1) to (3) and an iterative process, the interference order, N, of the maxima can be calculated and a refined value for refractive index for doped material can be obtained, in a similar way reported in ref. 35.

Values of n estimated for different structures (ribbon, sword and plate) are presented in figure 5, along with the n values for bulk pure ZnS obtained from equation (4), which agrees well with the value about 2.4 for visible range reported in the literature[36]. It can be seen that the analysis of resonant modes in the structures give reliable values for the refractive index for a wide range of wavelengths. In the present case of In doped ZnS structures a refractive index higher than that of the pure material has been observed. This refractive index increase could be due to the change in the crystal lattice parameter of ZnS associated to the incorporation of In. The ionic radius of $In^{3+}$ is larger than that of $Zn^{2+}$ ($r(In^{3+})$ = 81 pm vs. $r(Zn^{2+})$ = 74 pm), causing a slight increase in lattice parameter (as described in ref. 32), which is known be responsible for an increase of the refractive index of the material[37].

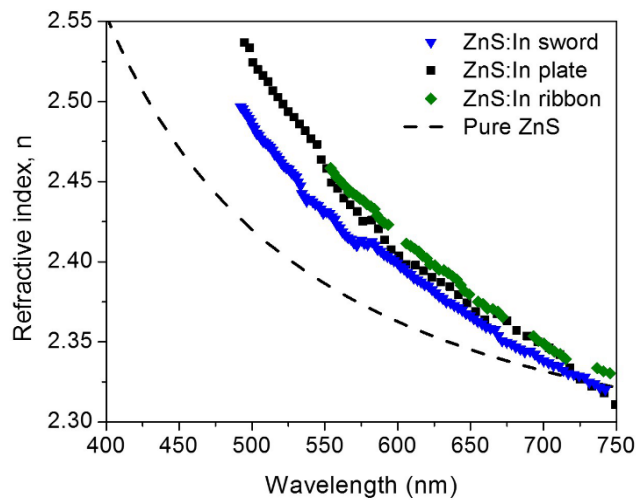

Figure 5. Experimental n values derived from the position of resonant maxima for each type of structure.

In the case of ZnS:Ga structures (pencils and rods) and ZnS:In plates, the results are not consistent with the existence of FPM, thus Whispering Gallery Modes (WGM) have been explored as responsible for the observed resonances.

Whispering gallery modes are obtained when the light is confined inside an optical cavity by total internal reflection (TIR), hence the light describes a closed optical path inside the cavity. As mentioned, these modes have been observed in hexagonal plates of ZnS:In and in pencils (with hexagonal cross section) and in wires with triangular cross section of ZnS:Ga.

For hexagonal ZnS:In plates different resonant modes can be supported. Figure 6 shows schematically the different modes that can be supported in a hexagonal optical cavity[38]. First, FP modes can be obtained if the light is reflected between two parallel faces. WG modes are achieved when the light suffers TIR on every face of the cavity, whereas quasi-WG modes are those obtained when light suffers TIR on one of every two faces. The optical path in each case depends on the length "a" of the side of the cavity.

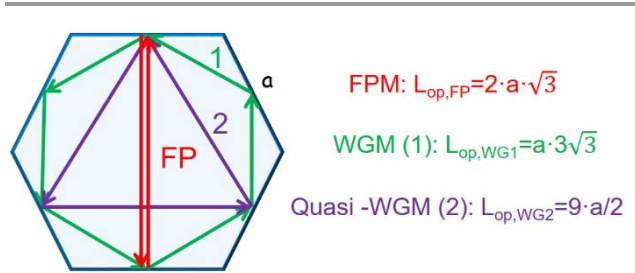

Figure 6. Different resonant modes that can be supported in a hexagonal cavity.

Figure 7 shows an example of WGM supported in a plate. In the SEM image (figure 7a) it can be seen that there are two stacked plates of different sizes, whereas the spatial distribution of the PL on these plates appears in figure 7b, where maxima of intensity are observed at the centre of the side faces of both plates, indicated with white arrows. The intense spot at the centre of the plate is the excitation point of the 325 nm light. The optical path inside the smaller plate is indicated in the schematic illustration of figure 7c, which corresponds to the WGM1 in figure 6.

ZnS:Ga structures have the PL emission centred in the blue-green region, at 510 nm, and can be associated with Ga-related defects as well as with native defects[23]. In this system, WGM have been observed in rods with hexagonal or triangular cross section. In the case of pencils (hexagonal rods) only the resonances associated to WGM1 are resolved. However a different situation is described for the triangular rods. In order to study the WGM in a triangular cavity, equations (5) to (8) are used for this type of cavity, taking into account that at least two different optical paths can be supported inside the cavity, as indicated in figure 8.

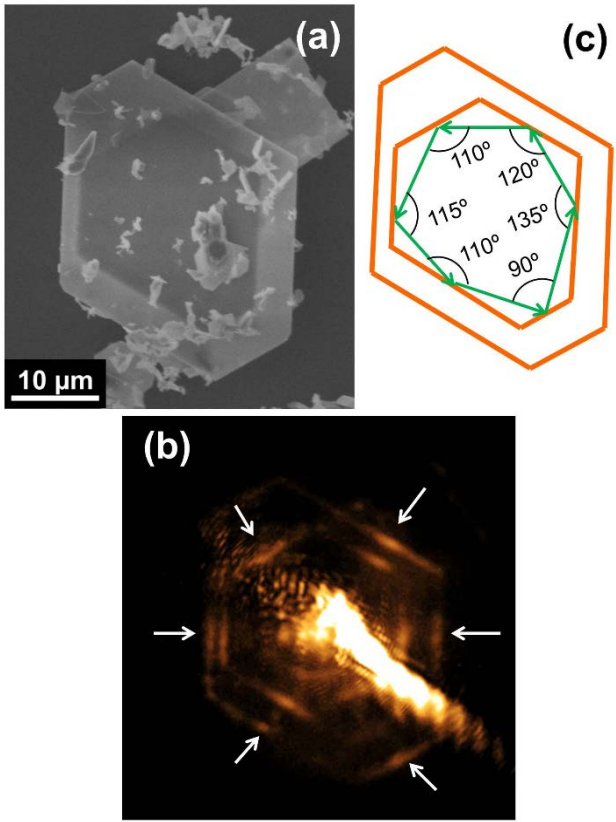

**Figure 7.** (a) SEM image of two stacking hexagonal plates of ZnS:In. (b) Optical image of the spatial distribution of the PL emission on the plate. (c) Optical path within the smaller plate.

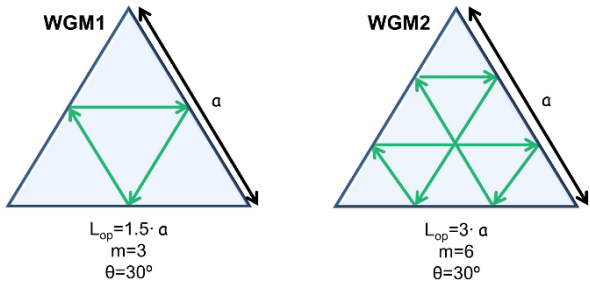

**Figure 8.** Different WGM that can be supported inside a triangular cavity.

In WGM, the position of the interference maxima depends on the phase shift occurring at TIR, which is different for the components of light perpendicular (⊥) and parallel (∥) to the plane of incidence. The phase shift (Φ) in each case is expressed by the following formulas[39]:

$$\phi_\parallel = -2\,arctan\left(n^2\sqrt{\frac{sen^2\theta - \frac{1}{n^2}}{1 - sen^2\theta}}\right) + \pi \quad (5)$$

$$\phi_\perp = -2\,arctan\left(\sqrt{\frac{sen^2\theta - \frac{1}{n^2}}{1 - sen^2\theta}}\right) \quad (6)$$

where θ is the incidence angle and n the refractive index of the material inside the cavity, and the WGM resonance wavelength for each polarization may be written as:

$$\lambda_\parallel = \frac{n \cdot L_{op}}{\left(N - \frac{m}{2}\right) + \frac{m}{\pi}arctan\left(n^2\sqrt{\frac{sen^2\theta - \frac{1}{n^2}}{1 - sen^2\theta}}\right)} \quad (7)$$

$$\lambda_\perp = \frac{n \cdot L_{op}}{N + \frac{m}{\pi}arctan\left(\sqrt{\frac{sen^2\theta - \frac{1}{n^2}}{1 - sen^2\theta}}\right)} \quad (8)$$

In this case 'm' is the number of TIR that light suffers inside the cavity and $L_{op}$ is the optical path. The shape of the cavity will determine the values of $L_{op}$, m and θ.

Figure 9a shows the μ-PL spectrum for the equilateral triangular section rod of ZnS:Ga, shown in the SEM image of figure 9b, (the position of the triangular cavity inside the wire, the axis of the structure and the polarization of the exciting laser are marked). Since both WGM1 and WGM2 are supported, the PL spectrum of figure 9a, shows a series of unevenly distributed resonant peaks corresponding to superposition of both modes. The modes can be separated by obtaining the linearly polarized spectra along the parallel and perpendicular direction to the cavity, as shown in figure 10.

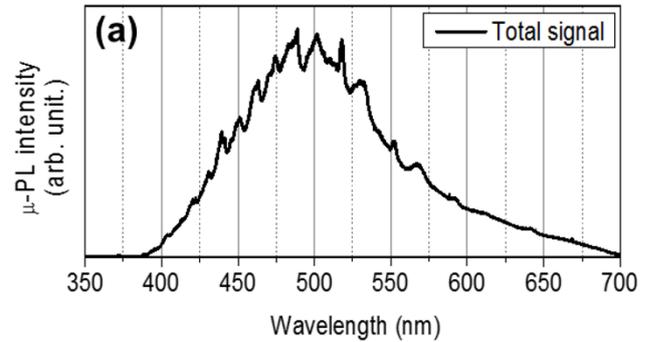

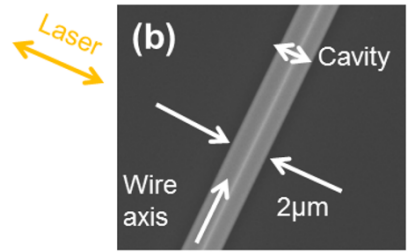

**Figure 9.** μ-PL spectrum of a ZnS:Ga wire with triangular cross section. (b) SEM image of the wire where the axis of the structure, the plane of the cavity and the polarization of the incident laser is indicated.

The equations for each polarization are:

**WGM1**

$$\lambda_\parallel = \frac{n \cdot L_{op}}{\left(N - \frac{3}{2}\right) + \frac{3}{\pi} arctan\left(n\sqrt{\frac{n^2-4}{3}}\right)} \quad (9)$$

$$\lambda_\perp = \frac{n \cdot L_{op}}{N + \frac{3}{\pi} arctan\left(\frac{1}{n}\sqrt{\frac{n^2-4}{3}}\right)} \quad (10)$$

**WGM2**

$$\lambda_\parallel = \frac{n \cdot L_{op}}{(N - 3) + \frac{6}{\pi} arctan\left(n\sqrt{\frac{n^2-4}{3}}\right)} \quad (11)$$

$$\lambda_\perp = \frac{n \cdot L_{op}}{N + \frac{6}{\pi} arctan\left(\frac{1}{n}\sqrt{\frac{n^2-4}{3}}\right)} \quad (12)$$

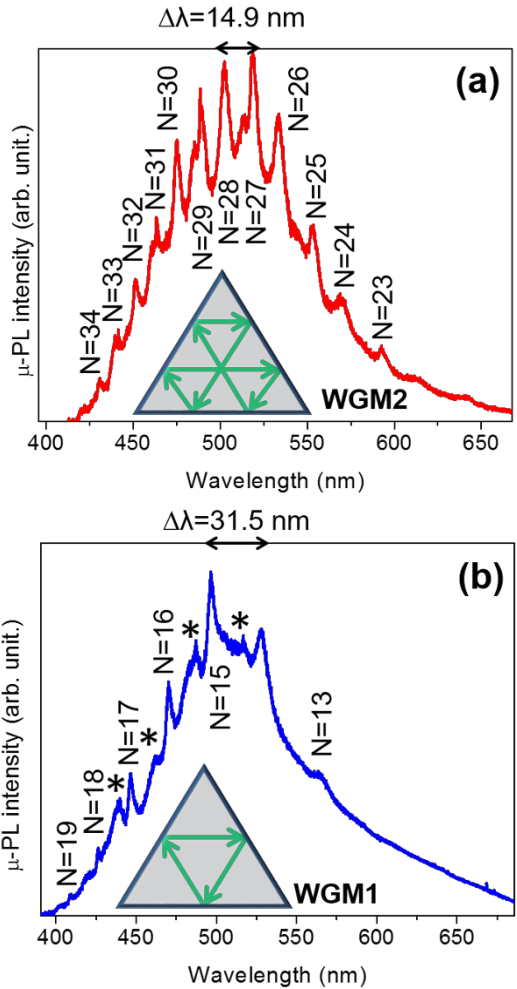

**Figure 10.** (a) µ-PL spectrum linearly polarized in the perpendicular direction to the triangular resonant cavity, i.e. parallel to the rod axis. (b) µ-PL spectrum linearly polarized in the parallel direction to the triangular resonant cavity, i.e. perpendicular to the rod axis.

For perpendicular modes, figure 10a, the separation between adjacent resonant peaks at the emission maximum is Δλ~14.9 nm, which for λ= 510 nm and n (510 nm) = 2.4, corresponds to an optical path Lop, of about 7 µm, which fits well to the WGM2 (figure 8a). Using equation (12), interference orders and refractive index can be calculated with the iterative process already described for FPM. N values are indicated in figure 10a, whereas the dispersion relation for the refractive index is presented in figure 11.

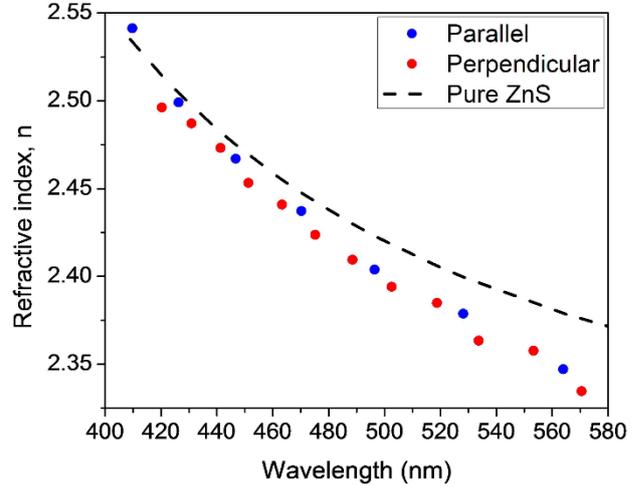

**Figure 11.** Experimental n values obtained from the position of resonant maxima for parallel and perpendicular modes.

On the other hand, parallel modes are presented in figure 10b. In this case, two series of peaks are observed (those marked with a number and those indicated with an asterisk). The separation between two adjacent peaks of the N series is Δλ~31.5 nm, which for λ= 510 nm and n(510 nm) = 2.4 fits well to the WGM1 with an optical path of 3.5 µm (figure 8). From equation (9), the interference orders N (figure 10b) and refractive index (figure 11) are estimated. The second series of peaks, with lower intensity, can be associated with the odd orders of WGM2, while even orders are overlapped with WGM1.

From these results, it can be seen that WGM1 are mostly polarized in the direction parallel to the cavity, while WGM2 are polarized in the perpendicular direction. This difference may be associated with different losses of each mode in the TIR on the cavity walls. Perpendicular modes would have lower losses, making them dominate in WGM2, where a higher number of reflections occur. These observations are consistent with the results obtained in other studies[40,41].

The refractive index for parallel and perpendicular modes are presented in figure 11. Data for pure ZnS are also shown for comparison. Parallel and perpendicular refractive index values are similar and slightly lower than those of pure material. This effect can be associated with a decrease of the lattice parameter[31], due to the smaller radius of $Ga^{3+}$ as compared with $Zn^{2+}$ radius, (r($Ga^{3+}$) = 62 pm vs. r($Zn^{2+}$) = 74 pm), which produces the opposite effect of $In^{3+}$ incorporation. Furthermore, a dependence of the refractive index with the direction of polarization of the confined light is observed. Wires with triangular cross section have the [0001] direction (c axis) contained in the plane of the resonant cavity[31]. Thus, the modes whose polarization is parallel to the c axis (i.e. parallel modes) have a higher refractive index, as described previously for ZnS[36].

Both curves can be fitted to the Cauchy's equation ($n=A+B/\lambda^2$), with A and B values presented in table 1.

|  | A | B (nm$^2$) | R$^2$ |
|---|---|---|---|
| **Perpendicular** | 2.147±0.005 | (63±1)·10$^3$ | 0.995 |
| **Parallel** | 2.135±0.009 | (67±2)·10$^3$ | 0.995 |

**Table 1.** Fitted parameters to the Cauchy's equation from data plotted in figure 11.

Finally, the dependence of PL emission on incident power has also been investigated. In figure 12a different μ-PL spectra for increasing laser power are presented. It can be seen that for increasing power, the intensity of the resonant peaks increases and their width decreases, which could be associated with a laser effect. For the highest power (figure 12b) two interesting peaks appear at 493 and 523 nm, with FWHM of 1.8 and 2.2 nm, respectively. To characterize the resonant cavity for these peaks, the quality factor, Q, and the finesse, F, can be calculated[35]:

$$Q = \frac{\lambda_{max}}{\Gamma} \quad (13)$$

$$F = \frac{\Delta\lambda}{\Gamma} \quad (14)$$

Where Γ stands for the full-width at the half maximum of the peak at λ. For the peaks at 493 and 523 nm, Q has respectively the values of 274 and 238 and F the values of 7.8 and 6.7. The Q values, close to 300, are high enough to use these resonant cavities as sensors or optical filters[42], however, the finesse values are lower than those reported for pure ZnS in the UV range[25,26]. Since the F factor is related to the losses produced at the walls of the cavity, (due to surface imperfections or absorption and scattering inside the cavity[41]), any improvement of the surface and hence reflectivity conditions would increase the quality of the resonant cavity.

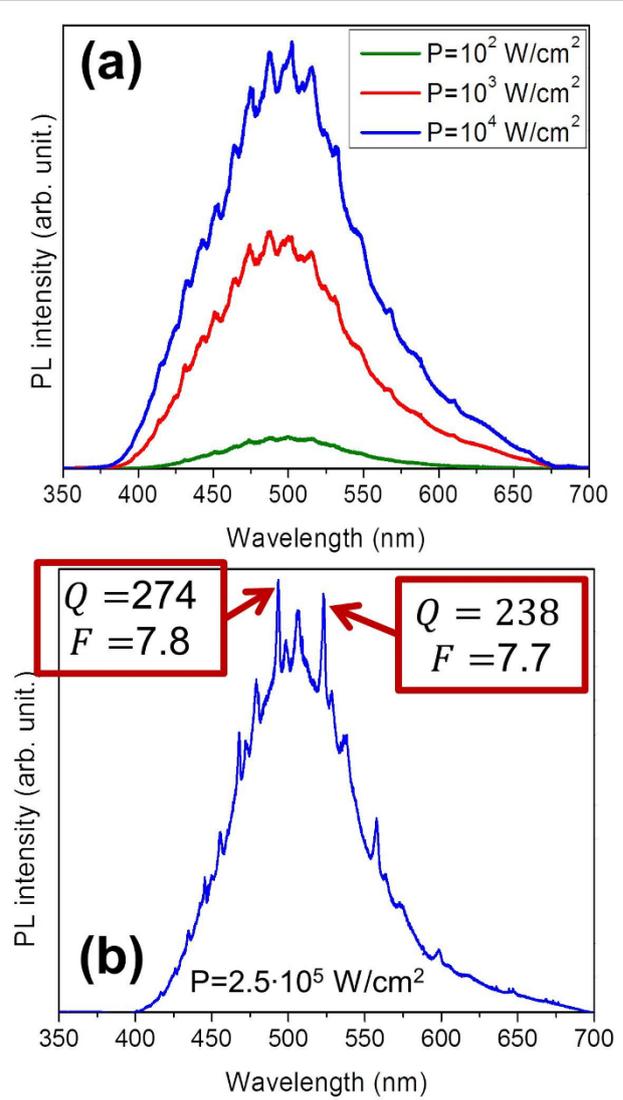

**Figure 12.** (a) Evolution of the μ-PL spectra of the ZnS:Ga wire with the laser power. (b) μ-PL spectrum for the highest laser power used. Q and F factors have been calculated for two resonant peaks at 493 nm and 523 nm.

## Conclusions

In this work we have studied resonant modes in optical cavities based on ZnS microstrcutures. In ZnS:In structures, FPM have been observed in different optical paths of ribbons, swords and nanoplates. In ZnS:In plates and ZnS:Ga rods with triangular or hexagonal cross sections, optical resonances are found to be WGM. The position of the resonant peaks and the values of the refractive index show a good agreement of the experimental results with the theoretical model for FPM and WGM. The ZnS:Ga wires could be good candidates to act as resonant cavities for applications as sensors or optical filters. The results obtained in ZnS:Ga wires with triangular section show that modes with higher number of reflections are mainly polarized in a direction perpendicular to the cavity, while the modes with fewer reflections are mainly polarized in the parallel direction, this indicates that the reflectivity is higher for the perpendicular components than for the parallel ones, which opens the possibility to improve the resonant cavity properties (in particular finesse). Laser effect in ZnS:Ga has been also observed.

## Acknowledgements

This work has been supported by MINECO (Projects MAT 2012-31959 and CSD2009-00013). B. Sotillo acknowledges Ministerio de Educación (Subprograma FPU) of Spain for financial support.

## References

1. X. S. Fang, Y. Bando, G. Z. Shen, C. H. Ye, U. K. Gautam, P. M. F. J. Costa, C. Y. Zhi, C. C. Tang and D. Golberg, *Adv. Mater.*, 2007, **19**, 2593-2596.
2. Z.-G. Chen, L. Cheng, H.-Y. Xu, J.-Z. Liu, J. Zou, T. Sekiguchi, G. Q. M. Lu and H.-M. Cheng, *Adv. Mater.*, 2010, **22**, 2376-2380.
3. F. Lu, W. Cai, Y. Zhang, Y. Li, F. Sun, S. H. Heo and S. O. Cho, *Appl. Phys.Lett.*, 2006, **89**, 231928.
4. D. Chen, F. Huang, G. Ren, D. Li, M. Zheng, Y.Wang and Z. Lin, *Nanoscale*, 2010, **2**, 2062-2064.
5. M.-Y. Lu, M.-P. Lu, Y.-A. Chung, M.-J. Chen, Z. L. Wang and L.-J. Chen, *J. Phys. Chem. C*, 2009, **113**, 12878-12882.
6. Q. Zhao, Y. Xie, Z. Zhang and X. Bai, C*ryst. Growth Des.*, 2007, **7**, 153-158.
7. X. Fang, Y. Bando, M. Liao, T. Zhai, U. K. Gautam, L. Li, Y. Koide and D. Golberg, *Adv. Funct. Mater.*, 2010, **20**, 500-508.
8. R. Xing, Y. Xue, X. Liu, B. Liu, B. Miao, W. Kang and S. Liu, *CrystEngComm*, 2012, **14**, 8044-8048.
9. J. H. He, Y. Y. Zhang, J. Liu, D. Moore, G. Bao and Z. L. Wang, *J. Phys. Chem. C*, 2007, **111**, 12152-12156.
10. K. Manzoor, S. R. Vadera, N. Kumar and T. R. N. Kutty, *Appl. Phys. Lett.*, 2004, **84**, 284-286.
11. M.-Y. Lu, J. Song, M.-P. Lu, C.-Y. Lee, L.-J. Chen and Z. L. Wang, *ACS Nano*, 2009, **3**, 357-362.
12. X. Fang, T. Zhai, U. K. Gautam, L. Li, L. Wu, Y. Bando and D. Golberg, *Prog. Mater. Sci.*, 2011, **56**, 175-287.
13. B. Y. Geng, L. D. Zhang, G. Z. Wang, T. Xie, Y. G. Zhang and G.W. Meng, *Appl. Phys. Lett.*, 2004, **84**, 2157-2159.
14. S. Biswas, S. Kar and S. Chaudhuri, *J. Phys. Chem. B*, 2005, **109**, 17526-17530.
15. J. Cao, D. Han, B. Wang, L. Fan, H. Fu, M. Wei, B. Feng, X. Liu and J. Yang, *J. Solid State Chem.*, 2013, **200**, 317-322.
16. A. Datta, S. K. Panda and S. Chaudhuri, *J. Solid State Chem.*, 2008, **181**, 2332-2337.
17. S. Kumar and N. Verma, *J. Mater. Sci. Mater. El.*, 2014, **25**, 785-790.
18. P. Yang, M. Lü, D. Xü, D. Yuan, J. Chang, G. Zhou, M. Pan, *Appl. Phys. A*, 2002, **74**, 257-259.
19. S. Lee, Y. Shin, Y. Kim, S. Kim and S. Ju, *J. Lumin.*, 2011, **131**, 1336-1339.
20. B. Cheng and Z. Wang, *Adv. Funct. Mater.*, 2005, **15**, 1883-1890.
21. C. C. Klick and J. H. Schulman, *Solid State Phys.*, 1957, **5**, 97-172.
22. B. Sotillo, P. Fernández and J. Piqueras, *J. Alloys Compd.*, 2014, **603**, 57-64.
23. B. Sotillo, P. Fernández and J. Piqueras, *J. Mater. Sci.*, 2015, **50**, 2103-2112.
24. B. Sotillo, Y. Ortega, P. Fernández and J. Piqueras, *CrystEngComm*, 2013, 15, 7080-7088.
25. J. A. Zapien, Y. Jiang, X. M. Meng, W. Chen, F. C. K. Au, Y. Lifshitz and S. T. Lee, *Appl. Phys. Lett.*, 2004, **84**, 1189-1191.
26. J. X. Ding, J. Zapien, W. Chen, Y. Lifshitz, S.-T. Lee and X. Meng, *Appl. Phys. Lett.*, 2004, **85**, 2361-2363.
27. V. Mulloni and L. Pavesi, *Appl. Phys. Lett.*, 2000, **76**, 2523-2525.
28. F. Vollmer, D. Braun, A. Libchaber, M. Khoshsima, I. Teraoka and S. Arnold, *Appl. Phys. Lett.*, 2002, **80**, 4057-4059.
29. T. Barwicz, M. A. Popovic, P. Rakich, M. Watts, H. Haus, E. Ippen and H. Smith, *Opt. Express*, 2004, **12**, 1437-1442.
30. M. A. Popovic, T. Barwicz, M. R. Watts, P. T. Rakich, L. Socci, E. P. Ippen, F. X. Kärtner and H. I. Smith, *Opt. Lett.*, 2006, **31**, 2571-2573.
31. B. Sotillo, Y. Ortega, P. Fernández and J. Piqueras, *Materials Research Express*, 2015, **2**, 035902.
32. B. Sotillo, P. Fernández and J. Piqueras, *J. Alloys Compd.*, 2013, **563**, 113-118.
33. D. Wang, H. W. Seo, C.-C. Tin, M. J. Bozack, J. R. Williams, M. Park and Y. Tzeng, *J. Appl. Phys.*, 2006, **99**, 093112.
34. H. H. Li, *Journal of Physical and Chemical Reference Data*, 1984, **13**, 103-150.
35. J. Bartolomé, A. Cremades and J. Piqueras, *J. Mater. Chem. C*, 2013, **1**, 6790-6799.
36. T. M. Bieniewski and S. J. Czyzak, *J. Opt. Soc. Am.*, 1963, **53**, 496-497.
37. K. Vedam, T. A. Davis, *Phys. Rev.*, 1969, **181**, 1196-1201.
38. J. Liu, S. Lee, Y. H. Ahn, J.-Y. Park, K. H. Koh and K. H. Park, *Appl. Phys. Lett.,* 2008, **92**, 263102.
39. M. Born and E. Wolf, *Principles of Optics*. Cambridge University Press, 1980.
40. G. Dai, Y. Zhang, R. Liu, Q. Wan, Q. Zhang, A. Pan and B. Zou, *J. Appl. Phys.*, 2011, **110**, 033101.
41. I. S. Grudinin, V. S. Ilchenko and L. Maleki, *Phys. Rev. A*, 2006, **74**, 063806.
42. C. Zhang, C.-L- Zou, Y. Yan, C. Wei, J.-M. Cui, F.W. Sun, J. Yao and Y. S. Zhao, *Adv. Opt. Mater.*, 2013, **1**, 357-361.